\documentclass[titlepage,12pt]{utarticle}   

\usepackage{graphicx}
\usepackage{amsmath,bm,amsfonts,amssymb,array,calc,amsthm,rotating,cite}
\usepackage{epsf}

\DeclareFontFamily{U}{rsf}{}
\DeclareFontShape{U}{rsf}{m}{n}{<5> <6> rsfs5 <7> <8> <9> rsfs7 <10-> rsfs10}{}
\DeclareMathAlphabet\Scr{U}{rsf}{m}{n}

\newcommand{\de}{\delta}

\newcommand{\g}{\chi}
\newcommand{\rh}{\rho}

\newcommand{\ta}{\tau}

\newcommand{\De}{\Delta}

\newcommand{\Om}{\Omega}

\newcommand {\cA} {{\cal A}}

\newcommand {\cF} {{\cal F}}

\newcommand {\cH} {{\cal H}}

\newcommand {\cL} {{\cal L}}

\newcommand {\cP} {{\cal P}}

\newcommand {\cT} {{\cal T}}

\newcommand {\cW} {{\cal W}}

\newcommand {\bbC} {\mathbb{C}}

\newcommand {\bbZ} {\mathbb{Z}}

\newcommand{\ba}{{\bar a}}

\newcommand{\td}{\tilde}

\newcommand{\unit}{{1\!\!\!\!1\,}}

\newcommand{\dl} {\partial}

\newcommand{\der}[1]{\frac{\partial}{\partial {#1}}}

\newcommand {\bra}{\bigl\langle}
\newcommand {\ket}{\bigr\rangle}

\newcommand{\rarrow}{\rightarrow}

\newcommand{\beq}{\begin{equation}}
\newcommand{\eeq}{\end{equation}}
\newcommand{\bea}{\begin{eqnarray}}
\newcommand{\eea}{\end{eqnarray}}
\newcommand{\nn}{\nonumber}

\newcommand{\beql}[1]{\begin{eqnarray}\label{#1}}

\newcommand{\eeql}{\end{eqnarray}}

\newcommand{\bit}{\begin{itemize}}
\newcommand{\eit}{\end{itemize}}

\newcommand{\ben}{\begin{enumerate}}
\newcommand{\een}{\end{enumerate}}

\long\def\bec #1 \eec{\begin{center}#1\end{center}}

\newtheorem{theorem}{Theorem}
\newtheorem{define}{Definition}

\long\def\bth #1 \eth{\begin{theorem}#1~\end{theorem}}
\long\def\bdef #1 \edef{\begin{define}#1~\end{define}}
\long\def\bpr #1 \epr{~\\ \emph{Proof:}~ #1~$\Box$}

\long\def\del#1\enddel{}
\long\def\new#1\endnew{{\bf #1}}

\def\ll{\label}
\def\nn{\nonumber{}}

\newcommand{\tfor}{{\qquad \mathrm{for} \quad}}

\def\ie{{\it i.e.}} 
\def\eg{{\it e.g.}} 
\def\Eg{{\it E.g.}}

\def\fla{_{a}}
\def\flb{_{b}}
\def\flc{_{c}}

\def\tria{\Delta}
\def\flab{_{\bar a}}
\def\flbb{_{\bar b}}
\def\flcb{_{\bar c}}
\def\fldb{_{\bar d}}
\def\fleb{_{\bar e}}
\def\flfb{_{\bar f}}
\def\supers#1{^{(#1)}}
\def\bc#1#2{\supers{i_#1,i_#2}}
\def\ial{{i}}
\def\Thetat{\Theta}
\font\fourteenmi=cmmi10 scaled\magstep 2
\def\BigP{{\hbox{$\textfont1=\fourteenmi \wp$}}}

\begin{document}        
\preprint{
  CERN--PH--TH/2006-038\\
  DESY-06-024\\
  ZMP-HH/06-04\\
  {\tt hep-th/0603085}\\
}


\title{\vskip-1cm Instanton Geometry and
Quantum $\cA_\infty$ structure \\on the Elliptic Curve}

\author{
Manfred Herbst${}^{a}$, Wolfgang Lerche${}^b$, Dennis Nemeschansky${}^c$
} 
        \oneaddress{
     ${}^a$ DESY Theory Group\\
     Notkestr. 85,
     22603 Hamburg,
     Germany\\
     {\tt Manfred.Herbst@desy.de}\\{~}\\
     ${}^b$ Department of Physics, CERN\\
     Theory Division,
     CH-1211 Geneva 23,
     Switzerland\\
    {\tt Wolfgang.Lerche@cern.ch}\\{~}\\
     ${}^c$ Department of Physics and Astronomy\\
     University of Southern California Los Angeles,
     CA 90089-0484, USA\\
     {\tt dennisn@usc.edu}\\
    }

    \nobreak
\Abstract{
We first determine and then study the complete set of non-vanishing
$A$-model correlation functions associated with the ``long-diagonal
branes'' on the elliptic curve. We verify that they satisfy the
relevant $\cA_\infty$ consistency relations at both classical and
quantum levels. In particular we find that the $\cA_\infty$ relation
for the annulus provides a reconstruction of annulus instantons out
of disk instantons. We note in passing that the naive application
of the Cardy-constraint does not hold for our correlators, confirming
expectations.  Moreover, we analyze various analytical properties
of the correlators, including instanton flops and the mixing of
correlators with different numbers of legs under monodromy. The classical
and quantum $\cA_\infty$ relations turn out to be compatible with
such homotopy transformations.  They lead to a non-invariance of
the effective action under modular transformations, unless compensated
by suitable contact terms which amount to redefinitions of the
tachyon fields. }

\date{March 2006}
\maketitle
\tableofcontents
\pagebreak

\section{Introduction}

By now, topological open string amplitudes, which determine important
terms such as the superpotential in the low-energy effective action,
have been well understood for single or multiple parallel $D$-branes.
However, more general configurations, such as ones described by
quiver diagrams based on intersecting branes, have not yet been
investigated in comparable detail, despite their potential importance
in phenomenological applications (see e.g.\ \cite{Blumenhagenmu}
for an overview).

For such configurations, the underlying TFT has a much richer
structure, which is due to the boundary changing operators that
describe open strings localized on intersecting $D$-branes. In fact,
the sophisticated mathematical machinery of homological mirror
symmetry \cite{kontsevich,mirrbook}, which acts between certain
categories of $A$- and $B$-type D-branes, gears up to full power
only for this kind of geometries. However, so far there have been
few applications in physics that make use of this structure, which
involves new types of open string instantons, or Gromov-Witten
invariants.

Indeed just a few explicit computations of boundary changing
correlation functions have been presented so far (we mean here
moduli-dependent, exact TFT correlators). One reason is that only
two- or three-point functions can be easily computed; for example,
for $B$-type branes via wavefunction overlaps \cite{Cremadeswa} or
 boundary LG models based on matrix factorizations \cite{Brunnermt}.
 Correlators for $A$-type branes can then be obtained from this via
 mirror symmetry, or simply by directly summing up instantons \cite
 {Cremadesqj}.

On the other hand, just like in the bulk theory, correlators with
more than three fields are difficult to evaluate directly, because
of the presence of integrated insertions which lead to singularities
that need to be regularized and may lead to contact terms. In the
bulk sector, however, the situation is favorable in that the moduli
space is flat and is governed by an integrable special geometry.
This implies that all higher-point correlators can be obtained as
derivatives from a generating function, the prepotential $\cF(t)$
\cite{DVV}. The contact terms are implicitly  determined by requiring
the vanishing of the Gau\ss-Manin connection \cite{Lossev}.
One can rephrase this also in terms of the WDVV equations which are
imposed as differential equations on~$\cF(t)$.

The situation is far more involved for the boundary sector, where
in general the moduli space is obstructed and there
is no notion of flatness (and correspondingly no a priori preferred
coordinates). The r\^ole of the bulk WDVV equations is replaced by
a set of generalized, open-closed WDVV equations \cite{HLL}, which
follow from various factorization and sewing constraints of
world-sheets with boundaries.  The simplest ones take the form of
$\cA_\infty$ relations \cite{Kajiurasn} between disk correlators;
 in the presence of bulk deformations, there are certain other
 conditions, following from bulk-boundary crossing symmetry and the
 factorization of the annulus amplitude. Very recently, a generalization
 of the $\cA_\infty$ relations to general Riemann surfaces with $h$
 boundaries and $g$ holes have been formulated in ref.~\cite{Herbstkt}
 and dubbed ``quantum $\cA_\infty$ relations''.

The purpose of this note is to gain insight in the interplay of the
various quantum $\cA_\infty$ relations, instanton sums and regularization
ambiguities, by considering $A$-model correlators for the simplest
brane geometry with moduli, namely for certain $D1$-branes on the
elliptic curve. The point is, of course, that the elliptic curve being flat is
very simple, and indeed one can solve the $\cA_\infty$ relations
in a completely geometric manner and determine the correlation
functions (as well as their ambiguities) in terms of instanton sums.
That is, correlators involving $N$ boundary changing operators can
be obtained by summing over the (moduli dependent) areas of $N$-gons,
 schematically:
$$
C_{a_1,...,a_N}(\tau)\ =\ \sum_{N-{\rm gons}} e^{-Area(\tau)}
$$
which correspond to world-sheet instantons whose boundaries lie on
the intersecting branes under consideration.\footnote{We will denote
the closed string modulus, \ie, the complexified K\"ahler parameter of
the curve, by $\tau$. The open string moduli, which correspond to
brane positions 
and Wilson lines and which are suppressed in the above formula,
will be generically denoted by $u$.} This technique has been pioneered
in \cite{Polishchukdb,PolishchukMA,PolishchukHMS,Polishchukkx,Polishchukoo},
where it was used to prove the mirror symmetry between the Fukaya
category of Lagrangian submanifolds, and the derived category of
coherent sheaves on the elliptic curve.

More concretely, we will first determine the complete set of
non-vanishing correlators pertaining to the ``long-diagonal'' branes,
which have already been discussed from various perspectives in
refs.~\cite {horiinf,Brunnermt,Govindarajanim}. While the three-point
functions have been explicitly computed before in
refs.~\cite{Polishchukdb,Cremadesqj,Brunnermt,Cremadeswa} and generic four-point correlators discussed in
\cite{PolishchukMA,PolishchukAP,PolishchukHMS,Polishchukkx}, we will evaluate the
remaining non-vanishing, higher-point disk and annulus correlators
by instanton counting and verify consistency with the classical and
quantum $\cA_\infty$ constraints (for transversal as well as certain
non-transversal brane configurations).

Moreover, we will discuss 
the analytical properties of correlators in non-technical terms,
 most notably singularities, ``instanton flops'' and ``homotopy''
 regularization ambiguities, all of which we give a simple geometrical
 interpretation.  Such homotopies are induced as monodromies from
 moving branes around the curve, and lead to a non-trivial fibration
 of the $\cA_\infty$ structure over the open/closed string moduli space. The
 effective superpotential is thus modular only up to homotopies,
 which may be compensated by simultaneous field redefinitions of the
 tachyons, or equivalently, by adding suitable contact terms.
We will also verify that homotopy transformations are compatible
with both the classical and the quantum $\cA_\infty$ constraints.

One of the most interesting results of this note concerns the annulus
quantum $\cA_\infty$ relation, for which we show that it maps
certain disk instantons to annulus instantons, essentially by
patching up the latter in terms of the former. This is a 
specific feature of open string instantons as it requires the fusion
of boundaries.

Finally, in an appendix we address the question of whether the
Cardy-type factorization relation (which is different to the annulus
quantum $\cA_\infty$ relation) holds or not. Although the familiar
factorization of the annulus diagram into closed or open string
channels is one of the fundamental axioms of open string TFT
\cite{MooreSegal,Moore,CIL}, it strictly speaking needs to apply
only to correlators without integrated insertions \cite{HLL}. We
show, by providing a counter-example, that the Cardy-constraint does
indeed not hold for general cylinder correlators on the elliptic
curve.

We hope that our findings will be useful for the understanding of more
complicated $D$-brane geometries, notably ones on Calabi-Yau
threefolds. 
\goodbreak
\section{Disk instantons and tree-level correlators}

\subsection{Recapitulation: 3-point functions for long-diagonal branes}

We will focus on the $A$-model and consider certain
$D1$-branes wrapped around the homology cycles of the elliptic
curve, $\Sigma$.  More specifically, in order to make contact with
previous work \cite{Brunnermt}, we will consider a specific
triplet of branes $\cL_i$ with $RR$ charges, or wrapping numbers
$(n,m)$ given by
\bea
\ll{Ldef}
\cL_1\sim (2,1)\,,\ \cL_2\sim (-1,1)\,,\ \cL_3\sim (-1,-2)\ .
\eea
These branes are usually referred to as ``long-diagonals'', which
is self-explaining upon drawing the branes on the covering space
of $\Sigma$ (see Fig.\ref{fig:branes} below). Each brane $\cL_i$
intersects the other ones three times within a fundamental domain.
This means that every boundary changing, open string vertex operator
that maps between a given pair of branes, carries an index $a$ that
labels the specific intersection at which it is located. 
The analysis of the cohomology (in the mirror LG-orbifold model)
\cite{Brunnermt} reveals that for an open string mapping from a 
brane $\cL_i$ to a brane $\cL_j$ at the intersection $a$, there is a
fermionic operator with $R$-charge 
$q=1/3$, which we denote by $\Psi^{(i,j)}\fla$. Moreover, there is
a ``Serre dual'' bosonic operator $\Phi^{(j,i)}\flab$ of charge
$q=2/3$ for an open string going the other way. Finally, apart from
the identity operator, there are fermionic, boundary preserving
operators $\Omega^{(i,i)}$ of charge $q=1$, which are tied to single
branes $\cL_i$ and which are the marginal operators coupling to the
brane moduli, $u_i$.

The simplest correlators of these operators give the topological
open string metric:
\bea
  \ll{metric}
  \rh_{\unit\Om} &:=& \bra~ \unit^{(i,i)} \unit^{(i,i)} \Om^{(i,i)} \ket_{disk}
  = 1 \\
  \rh_{a\ba} &:=& \bra~ \unit\bc{1}{1} \Psi_a\bc{1}{2} \Phi_\ba\bc{2}{1}\ket_{disk}
  = \de_{a\ba} \ .\nonumber
\eea
On the other hand, the simplest non-trivial correlation functions are
the following three-point functions:
\bea
\label{threept}
\tria\supers{i_3i_1i_2}_{abc}(\tau,u_i)\ = \ 
\bra
    \Psi\fla\bc{3}{1}\Psi\flb\bc{1}{2}
       \Psi\flc\bc{2}{3}  
       \ket_{disk}\ ,
\eea
which have been evaluated in the $B$-model using wavefunction
overlaps in \cite{Cremadeswa} or using the LG model based on
matrix factorization \cite{Brunnermt}. The result, when expressed
in terms of the flat coordinates $\tau$, $u_i$ (which coincide with
the natural variables of the mirror $A$-model), looks:
\beq
  \label{triangle}
\tria\supers{i_3i_1i_2}_{abc}(\tau,u_i)\ =\ 
\delta_{a+b+c,0}^{(3)} \, 
\Thetat \left[\begin{array}{c} \![b-c]_3-3/2\! \\ -3/2 \end{array}\right]
\big(\,3\tau \,\vert\, 3(u_1+u_2+u_3)\,\big)\ ,
\eeq
where $\delta^{(3)}$ is the Kronecker function defined modulo three,
and $[a]_3$ denotes the mod $3$ reduction of $a\in\bbZ$ to the range
$\{1,2,3\}$. Furthermore,  
\beq  
\label{theta}
   \Thetat\left[
      \begin{array}{c} a \\ b\end{array} \right]
	\big(\,3\tau\,\vert\, 3u\,\big)
       = 
    \sum_{n = - \infty}^\infty 
      q^{\frac16(a+3n)^2} e^{2 \pi i (u+b/9)(a+3n)} \ ,
\eeq
is a standard theta-function whose expansion in $q\equiv e^{2\pi
i\tau}$ sums up the contributions of all the triangular world-sheet
instantons that are bounded by the three branes $\cL_i$ (with moduli
$u_i$).\footnote{To properly describe the $u$-dependence of the
areas, one would need to multiply these correlators with simple
non-holomorphic ``quantum'' prefactors \cite{Cremadeswa}, which may be viewed as arising
from a holomorphic anomaly (indeed these can be traced back to the
holomorphic anomaly of the annulus amplitude).  However, since TFT
computations naturally yield holomorphic expressions, we suppress
such factors here. They could be easily reinstated by requiring
modular covariance, if one wished to do so.} This is
completely in line with the findings of ref.~\cite{Polishchukdb}. We visualize
the situation in Fig.\ref{fig:branes}, which also serves to define our conventions for labeling branes and boundary fields.

\begin{figure}[t]  
\begin{center}
\includegraphics[width=10.5cm]{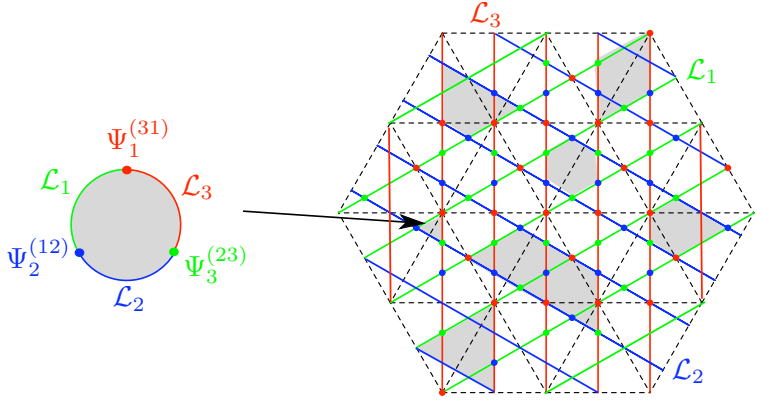}
\caption{On the right we have drawn the long-diagonal branes $\cL_{1,2,3}$
on the covering space of the elliptic curve, as well as some examples
of polygon instantons  bounded by  these branes. On the left we have
drawn a world-sheet disk with three boundary changing operator
insertions, which gets mapped into certain triangle shaped instantons
on the right. In our conventions, red, blue and green dots correspond
to operators $\Psi^{(i,j)}_a$ for $a=1,2,3$, respectively. Moreover, the shown
brane locations correspond to our choice of origin in the open string
moduli space. 
}
\label{fig:branes}
\end{center}
\end{figure}

\subsection{Polygon instantons and higher-point amplitudes}

The three-point functions (\ref{threept}) are only the first terms
of the effective superpotential, and one of our purposes is to
determine the complete superpotential that can be associated with
the three types of branes $\cL_i$ of (\ref{Ldef}). Due to $R$-charge
selection rules, there is only a finite number of terms, and in our
situation the maximal number of external ``tachyon'' legs is
$N=6$.\footnote{Allowing for general branes would involve an infinite
number of terms, because the charge of the boundary changing operators
is correlated with the angle of the intersecting branes, and this
can become arbitrary small for generic wrapping numbers $(m,n)$.}

Schematically, the effective superpotential can be written in the following form:
\bea
\label{weff}
\cW_{eff}(\tau,u_i,t_a,\xi_{\bar a}) &=& 
\frac 13 \tria_{abc}(\tau,u) t_at_bt_c
+\frac 12 \cP_{a\bar bc\bar d}(\tau,u)t_a\xi\flbb t_c\xi\flcb
+\cT_{ab\bar c\bar d}(\tau,u)t_at_b\xi\flcb\xi\fldb
\\[2mm]&+&
\BigP_{a\bar b\bar c\bar d\bar e}(\tau,u)t_a\xi\flbb\xi\flcb\xi\fldb\xi\fleb
+\frac 16 \cH_{\bar a\bar b\bar c\bar d\bar e\bar f}(\tau,u)\xi\flab\xi\flbb\xi\flcb\xi\fldb\xi\fleb\xi\flfb\ .\nn
\eea
Note that for a given brane configuration, not all terms may
contribute. (\Eg, if all D-branes wrap the same homology class then 
all terms vanish, reflecting that there is no obstruction to move the branes
around; if the D-branes wrap only two homology classes then only the
second term in the superpotential is non-trivial. We will see in a
moment that the latter is associated with parallelogram instantons.) 
Above, $t_a$ are the bosonic deformation parameters that
couple to the fermionic open string tachyons $\Psi_a$, while the
$\xi\flab$ are fermionic parameters coupling to the bosonic operators,
$\Phi_{\bar a}$.\footnote{One can view $(t,\xi)$ as coordinates of
a non-commutative superspace, see e.g., \cite{Lazaroiukm}.}  
Note that we suppressed the labels $i$ of the branes.

More specifically, the various inequivalent, cyclically 
symmetric disk correlators with $N\geq4$ legs are defined as follows:
\bea
  \ll{disk}
  \cT_{a b \bar c \bar d}^{(i_4i_1i_2i_3)} &=& 
      \bra \Psi\fla\bc{4}{1}\Psi\flb\bc{1}{2}
      \Phi\flcb\bc{2}{3}\Phi\fldb\bc{3}{4}
      \ket_{disk} \\
  \cP_{a \bar b c \bar d}^{(i_4i_1i_2i_3)} &=& 
      \bra \Psi\fla\bc{4}{1}\Phi\flbb\bc{1}{2}
      \Psi\flc\bc{2}{3}\Phi\fldb\bc{3}{4}
      \ket_{disk}\nn \\
     \BigP_{a\bar b\bar c\bar d\bar e}^{(i_5i_1i_2i_3i_4)} &=& 
      \bra \Psi\fla\bc{5}{1}\Phi\flbb\bc{1}{2}
      \Phi\flcb\bc{2}{3}\Phi\fldb\bc{3}{4}\Phi\fleb\bc{4}{5}
      \ket_{disk}\nn \\
  \cH_{\bar a\bar b\bar c\bar d\bar e\bar f}^{(i_6i_1i_2i_3i_4i_5)} &=& 
      \bra \Phi\flab\bc{6}{1}\Phi\flbb\bc{1}{2}
      \Phi\flcb\bc{2}{3}\Phi\fldb\bc{3}{4}
         \Phi\fleb\bc{4}{5}\Phi\flfb\bc{5}{6}
      \ket_{disk} \nn
      \ ,
\eea
where it is implicitly understood that $(N-3)$ operators are
integrated as topological descendants. 

Correlators with $N$
boundary changing insertions will generically get contributions of 
world-sheet instantons that end on $N$ intersecting branes $\cL_i$,
which thus can be depicted as $N$-gons on the covering space of the
curve; there are two different geometries for $N=4$, namely given by
trapezoids ($\cT$) and parallelograms ($\cP$). Since two branes can
always be fixed using translational invariance, $N$-point correlators involving $N$ branes will
depend on only $N-2$ independent combinations of the brane moduli $u_i$. 
Note, moreover, that the angles $\varphi$ at the corners of an $N$-gon
are related to the $R$-charge $q$ of the boundary changing field, \ie,
$\varphi=q \pi$.

In addition to the fields in the boundary changing sectors, we
allow for an arbitrary number of marginal operator
insertions, $\Om^{(i,i)}$.  Since these are associated with the
flat coordinates $u_i$ \cite{Brunnermt} which are integrable, we can
obtain such correlators simply by taking partial derivatives
$\dl_{u_i} \equiv\der{u_i}$, e.g.,
\[
  \frac{1}{(6 \pi i)^{n}}
  \dl^{n_1}_{u_{i_1}} \dl^{n_2}_{u_{i_2}} \dl^{n_3}_{u_{i_3}} 
  \De^{(i_3i_1i_2)}_{abc} = 
    \bra
      \Psi\fla\bc{3}{1} \left(\Om\bc{1}{1}\right)^{n_1}
      \Psi\flb\bc{1}{2} \left(\Om\bc{2}{2}\right)^{n_2}
      \Psi\flc\bc{2}{3} \left(\Om\bc{3}{3}\right)^{n_3}
    \ket_{disk} \ ,
\]
where $n = n_1 + n_2 + n_3$. This readily generalizes to all the other
amplitudes in (\ref{disk}). 

The evaluation of the correlators (\ref{disk}) proceeds by identifying
the smallest $N$-gon on the covering space that can be associated
with the given boundary conditions and determining the area of it as
well as of all other $N$-gons obtained by lattice translations from
it. Summing all areas up we count instantons
and anti-instantons with opposite orientations.
The result will have the form of a generalized, indefinite
theta-function \cite{Polishchukoo};  this will be a section of
some vector bundle over $\Sigma$, much like the ordinary theta-function
(\ref{theta}) is a section of a line bundle, $L^{\otimes3}$.

For example, let us consider a trapezoid associated with the
correlator $\cT_{a b \bar c \bar d}$, and label the boundary fields
in a manner as shown in Fig.\ref{fig:branes}.  The shorter of the
two parallel sides then has length $l_{short}=[\bar d-\bar c+3/2]_3+3m$,
where $m\in \bbZ$ accounts for lattice shifts.  Similarly, the two
sides of equal length have $l_{diag}=[b-\bar c]_3+3n$, and the
longer side has $l_{long}=l_{short}+l_{diag}$. The area of the
trapezoid is thus  $A= 1/2([b-\bar c]_3+3n)(2[\bar d-\bar
c+3/2]_3+[b-\bar c]_3+3n+6m)$. All-in-all, when allowing for
continuous translations parametrized by $u_i$,\footnote{Note that
this expression and analogous ones discussed below are defined only
for appropriate $u_i$; we will address this issue in the next
section. Also note, just as for the three-point function, that in
order to describe the correct area dependence on the $u_i$, we would
need to add a simple non-holomorphic prefactor that we suppress
here.} we obtain:
\bea
\ll{trapezoid}
\cT_{a b \bar c \bar d}^{(i_4i_1i_2i_3)}(\tau,u_i) \ =\
\delta_{a+b,\bar c+\bar d}^{(3)} \,
   \Theta_{trap}\left[
    \begin{array}{c} [b-\bar c]_3 \\[0pt] [\bar d-\bar c+3/2]_3 \end{array}
    \right] (3\tau | 3(u_1+u_2+u_4),3(u_1-u_3) )\ ,
\eea
where the trapezoidal theta-function is defined by the following
indefinite series:
\bea
 \label{deftrap}
  \Theta_{trap}\left[\begin{array}{c}a \\ b\end{array}\right] 
  (3\tau| 3u,3v) \ =\
  \sum_{m, n \in \bbZ}^{\mathrm{indef.}}
  q^{\frac16 (a + 3n)(a  + 3n + 2(b+3m))} 
  e^{2\pi i \bigl( 
      (a + 3n) (u-1/6) +
      (b + 3m) v
    \bigr)}\ ,
\eea
with 
\bea
\label{indefsum}
\sum_{m,n \in \bbZ}^{\mathrm{indef.}} \ \equiv\ \sum_{m,n=0}^{\infty} - \sum_{m,n=-1}^{-\infty}.
\eea
In a similar way, one finds for the parallelogram correlators:
\bea
 \label{parallelogram}
   \cP_{a \bar b c \bar d}^{(i_4i_1i_2i_3)}(\tau, u_i)\ &=&\
  \delta^{(3)}_{a+c,\bar b+\bar d}~
  \Theta_{para}\left[
    \begin{array}{c} [c-\bar b]_3 \\[0pt] [\bar d-c]_3\end{array}
    \right] (3\tau | 3(u_1-u_3),3(u_4-u_2)) \ ,
    \\
  \Theta_{para}\left[\begin{array}{c}a \\ b\end{array}\right] 
  (3\tau  | 3u,3v) &\equiv&
  \sum_{m, n \in \bbZ}^{\mathrm{indef.}}
  q^{\frac13 (a + 3n)(b + 3m)} 
  e^{2\pi i \bigl( 
      (b + 3m) u +
      (a + 3n) v
    \bigr)} \nonumber    
    \ .
    \eea
    
The five-point function looks more difficult to determine, but we use a trick in order to make life simpler: there is one side of the pentagon that is not parallel to any other one - when we attach a triangle to it, the pentagon turns into a parallelogram. So we can describe the area of the pentagon as the difference of a parallelogram and a triangle (taking of course all lattice translations into account). Taking everything together, we obtain:
 \beq
 \label{fivepoint}
   \BigP_{a\bar b\bar c\bar d\bar e}^{(i_5i_1i_2i_3i_4)}(\tau, u_i) =
  \delta^{(3)}_{a,\bar b+\bar c+\bar d+\bar e}~
  \Theta_{penta}\!\left[\!\!
    \begin{array}{c} [-b\!-\!c\!-\!d]_3 \\[0pt] 
                     [e\!+\!c\!+\!d]_3 \\[0pt] 
		     [c\!-\!d\!+\!\frac 32]_3
    \end{array}
    \!\!\right]\! 
  (3\tau | 3(u_5\!-\!u_2),3(u_1\!-\!u_4),3(u_3\!+\!u_2\!+\!u_4)) \ , 
\eeq
where
\bea
  \nonumber
  \Theta_{penta} && \hspace{-20pt}\left[
    \begin{array}{c}a \\ b \\ c\end{array}
    \right]
  (3\tau| 3u,3v,3w) \equiv \\[5pt]
  \equiv &&\hspace{-15pt} \sum_{m, n, k \in \bbZ}^{\mathrm{indef.}}
  q^{\frac 13(a_{>}+3(n+k))(b_{>}+3(m+k)) - \frac 16 (c+3k)^2} 
  e^{2\pi i \bigl( 
               (a_{>}+3(n+k)) u +
               (b_{>}+3(m+k)) v +
               (c+3k) (w-1/6)
            \bigr)}    
  \nonumber  \ ,
\eea
where $a_> = a+3$ for $a<c$ and $a_> = a$ for $a>c$, and similarly for
$b_>$. The shifts in $a_>$ and $b_>$ ensure that the sides of the
parallelogram are longer than the side of the subtracted triangle.
The indefinite sum is defined as
\[
  \sum_{m, n, k \in \bbZ}^{\mathrm{indef.}} = 
  \sum_{m, n, k \geq 0}^\infty + \sum_{m, n, k \leq -1}^{-\infty}
\]

Finally, we find the six-point functions by subtracting two triangles
from the acute-angled corners of a parallelogram:
 \bea
 \label{sixpoint}
  \cH_{\bar a\bar b\bar c\bar d\bar e\bar f}^{(i_6i_1i_2i_3i_4i_5)}
  (\tau, u_i)
  &=& \delta^{(3)}_{0,\bar a + \bar b + \bar c + \bar d + \bar e +
  \bar f} 
  \times \\[5pt] \nn
  \times\Theta_{hexa} \left[\!\!
    \begin{array}{c} [-b\!-c\!-d]_3 \\[0pt] 
                     [c+\!d+\!e]_3\\[0pt] 
                     [c-\!d+\!\frac32]_3\\[0pt]
                     [a-\!f+\!\frac32]_3
    \end{array} \!\!\right]\hspace{-1cm}
    && (3\tau | 3(u_5\!-\!u_2),3(u_1\!-\!u_4),
        3(u_3\!+\!u_2\!+\!u_4),3(-\!u_6\!-\!u_1\!-\!u_5)) 
  \ ,
\eea
where
\bea
  \nn
  \Theta_{hexa} && \hspace{-20pt}
     \left[
       \begin{array}{c}a \\[-3pt] b\\[-3pt] c\\[-3pt] d
       \end{array}
     \right] 
  (3\tau | 3u,3v,3w,3z) \equiv \\[5pt]
  \equiv &&\hspace{-15pt} 
  \sum_{m, n, k, l \in \bbZ}^{\mathrm{indef.}}
  q^{\frac 13(a+3n)(b+3m) - \frac 16 (c+3k)^2 - \frac 16 (d+3l)^2} 
  e^{2\pi i \bigl( 
               (a+3n) u +
               (b+3m) v +
               (c+3k) (w-1/6) +
               (d+3l) (z+1/6)
            \bigr)}    
  \nonumber .
\eea
The indefinite sum is given by
\[
  \sum_{m, n, k, l \in \bbZ}^{\mathrm{indef.}} =
  \sum_{m, n \geq 0}^{\infty} 
  \sum_{k \geq 0}^{<k_{max}} 
  \sum_{l \geq 0}^{<l_{max}} -
  \sum_{m, n \leq -1}^{-\infty} 
  \sum_{k \leq -1}^{>k_{min}} 
  \sum_{l \leq -1}^{>l_{min}}
\]
with $k_{max}=\min(a/3+n,b/3+m)-c/3$ and 
$l_{max}=\min(a/3+n,b/3+m)-d/3$ as well as 
$k_{min}=\max(a/3+n,b/3+m)-c/3$ and
$l_{min}=\max(a/3+n,b/3+m)-d/3$. The restrictions in the sums ensure
that the subtracted triangles are not larger than the parallelogram.

We have so far defined all topological A-model disk amplitudes in terms of
instanton sums for the D-brane configuration as shown in
Fig.\ref{fig:branes}. In the following we proceed investigating their
analytic properties.

\subsection{Analytical properties of the disk amplitudes}

The correlation functions we wrote down in the previous section are
not completely well-defined. There are the following three inter-related
issues that we need to discuss: i) singularities from colliding
branes;  ii) analytic continuation over the open string moduli
space; iii) modular anomalies; and iv) contact term ambiguities
intrinsic to the definition of the correlators.  Most of these
aspects have been, in one form or the other, already discussed in the
mathematical literature (see
\cite{PolishchukAP,PolishchukHMS,PolishchukMA,Polishchukkx}),
although they have not yet been exhibited in the string physics
literature.  We found it instructive to work out, as a case study,
some of these aspects for our specific brane geometry, and in order to aid the non-expert, we will present them in simple non-technical terms.  We
will focus on the trapezoid correlator, as it captures the relevant
features, with the understanding that the other, higher-point
functions can be dealt with analogously.

One basic point is that the trapezoid sum (\ref{deftrap}), as well
as the other  higher-point functions, is defined as a sum over
instantons with positive areas; the summation (\ref{indefsum}) is
equivalent to requiring $(a + 3n)(a + 2b + 3n + 6m)>0$, \ie,
positive area. This assumes for the moduli that $0\leq  u_1,v_1< 1$
where $u\equiv: u_1\tau+u_2$, $v\equiv: v_1\tau+v_2$.  However,
when doing large reparametrizations, such as shifts $u\rightarrow
u-k\tau$, $v\rightarrow v-\ell\tau$, we may formally produce negative
areas and so leave the domain of support; this is similar to the
phenomenon of leaving the K\"ahler cone of a Calabi-Yau manifold.
We thus expect that some suitable analytic continuation, describing
the analog of a flop transition to a different geometry\footnote{Or
``different phase'' in the language of
ref.~\cite{Aganagicgs,Aganagicnx}, where similar phenomena
were considered for non-compact branes.} with positive instanton
areas, will be necessary.

To proceed, let us rewrite the trapezoid sum (\ref{deftrap}) in a
form similar to an Appell-function \cite{PolishchukAP} as follows:
\beq
\label{defkappa}
  \Theta_{trap}\left[\begin{array}{c}a \\ b\end{array}\right] 
  (3\tau | 3u,3v) \ =\ e^{2\pi i v b}  \sum_{n \in \bbZ}
 \frac{ q^{\frac16 (a + 3n)(a + 2b + 3n)} 
  e^{2\pi i 
      (a + 3n) (u-1/6) }}
      {1-q^{a + 3n}e^{6\pi i v}}\ .
\eeq 
This way of representing it makes the singularity manifest which
occurs when the parallel sides move on top of each other (and the
Wilson line is tuned appropriately), \ie, for $3v+\tau a\in\bbZ+3\tau\bbZ$.
The $1/(1-x)$ singularity results from summing infinitely many
instantons that degenerate to zero area, and signals the appearance
of extra physical states \cite{Govindarajanim}.

\begin{figure}[t]  
\begin{center}
\includegraphics[width=8cm]{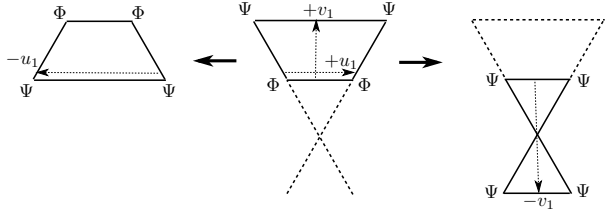}
\caption{Degenerations of the trapezoidal instanton. For positive
$u_1,v_1$ (middle figure), the area grows monotonically, while for
negative $u_1$ or $v_1$ we formally run into the depicted, partly
ill-defined geometries with negative areas. Resumming the instanton
series amounts to a flop-like transition back to positive area, albeit
for a different geometry.
}
\label{fig:trapdeg}
\end{center}
\end{figure}
However we can go on and further continue $v_1$ to negative values
(thereby avoiding the singularity by switching on $v_2$, \ie, a Wilson line),
by making use of the identity $\frac x{x-1}=-\frac1{(1/x)-1}$
in (\ref{defkappa}). We find:
\beq
\label{kappatransf}
  \Theta_{trap}\left[\begin{array}{c}a \\ b\end{array}\right] 
  (3\tau | 3u,-3v)
   \ =\ -e^{-2\pi ia/3}e^{6\pi iv}\,
   \Theta_{trap}\left[\begin{array}{c}-a \\ -b\end{array}\right] 
  (3\tau | 3(\tau-u),3v)
  \ .
\eeq 
This shows how resummation of the instanton series maps back to a
well-defined, however different geometry (the r\^oles
of the boundary fields can formally change, \ie, $\Psi$'s transmute
into $\Phi$'s and vice versa); one might call this phenomenon an
``instanton flop''. See Fig.\ref{fig:trapdeg} for a sketch of this.

More generally, from (\ref{defkappa}) we can deduce the following
behavior under shifts of the open string moduli:
\def\thf{{\frac32}}
\bea
\ll{periodicity}
  \Theta_{trap}\!\left[\begin{array}{c}a \\ b\end{array}\right]\! 
  (3\tau \,| \,3u,\,3v)  &=& 
  e^{\mp2\pi ia}\,\Theta_{trap}\!\left[\begin{array}{c}a \\ b\end{array}\right]\! 
  (3\tau \,| \,3(u\pm1),\,3v)
  \nonumber\\  &=& 
 e^{\mp2\pi ib} \, \Theta_{trap}\!\left[\begin{array}{c}a \\ b\end{array}\right]\! 
  (3\tau\, | \,3u,\,3(v\pm1))\ ,
   \\
     \Theta_{trap}\!\left[\begin{array}{c}a \\ b\end{array}\right]\! 
  (3\tau \,| \,3u,\,3(v\pm\tau)) &=&
  e^{\mp6\pi i(u-1/6)}q^{3/2}\,\Theta_{trap}\!\left[\begin{array}{c}a \\ b\end{array}\right]\! 
  (3\tau \,| \,3(u\mp\tau),\,3v)\ ,
 \nonumber \\
    \Theta_{trap}\!\left[\begin{array}{c}a \\ b\end{array}\right]\! 
   (3\tau \,| \,3(u\pm\tau),\,3v) &=&
  e^{\mp6\pi iv} \Theta_{trap}\!\left[\begin{array}{c}a \\ b\end{array}\right]\! 
   (3\tau \,| \,3u,\,3v) 
    \nonumber\\ &\mp&\!\!
 e^{-2\pi i(u-\frac16)(b-\thf\pm\thf)} e^{2\pi iv(b-\thf\mp\thf)}q^{-\frac16(b-\thf\pm\thf)^2} \Theta\!\left[\begin{array}{c}\!a\!+\!b\! \\ -3/2\end{array}\right]\! 
  (3\tau \,| \,3u).
  \nonumber
\eea
The ordinary theta function in the last equation may be viewed as
an anomaly or obstruction for the trapezoid function against being
(quasi-)periodic, and reflects that the Appell function, together
with $\Theta$, forms a section of a {\it non-split} rank two vector
bundle.  In physical terms, this simply means that the trapezoid
function does not extend nicely over the covering space, but rather
gets an extra contribution in the form of a three-point function
when we translate a brane around the torus.

The mixing of different correlation functions under monodromy
renders the effective superpotential (\ref{weff}) ambiguous and
non-modular.  One may remedy this by defining an invariant
correlator in the following manner:
\bea
\ll{newcorrel}
\bar \Theta_{trap}\!\left[\!\!\begin{array}{c}a \\ b\end{array}\!\!\right] 
 \! (3\tau | 3u,3v)\ \equiv\ 
   \Theta_{trap}\!\left[\!\!\begin{array}{c}a \\ b\end{array}\!\!\right] 
  (3\tau | 3u,3v) - P_{b}(q,u\!-\!v\!-\!1/6)\, 
   \Theta\! \left[\!\!\begin{array}{c}a+b \\
     -3/2\end{array}\!\!\right]\!(3\tau \,| \,3u),
\eea
where $P_{b}(x)$ is a piecewise polynomial function in $e^{2\pi
ix}$ that is designed \cite{PolishchukMA} to cancel the $\Theta$
function terms in (\ref{periodicity}). Explicitly, we find for our
geometry ($x_1\equiv\frac{{\rm Im}x}{{\rm Im}\tau}$): 
\bea
\ll{thedreadedP}
P_{b}(q,x)\ =\ 
\Bigg\{
\begin{array}{lll} 
 -q^{-\frac16b^2}\sum_{n=1}^m q^{-\frac32n^2+3mn+b(n-m)}
 e^{2\pi i(3n-b)x},&
m=[x_1] &   \textrm{for}\ x_1\geq0\\
q^{-\frac16(b-3)^2}\sum_{n=1}^{-m} q^{-\frac32n^2-3mn+(3-b)(n+m)}
e^{-2\pi i(3n+b-3)x},&
 m=-[-x_1] &   \textrm{for}\ x_1<0. \\
\end{array} 
\eea
The correlator (\ref{newcorrel}) then has indeed the desired global
properties over the full open string moduli space, \ie, is
(quasi-)periodic:
\bea
\bar \Theta_{trap}\left[\begin{array}{c}a \\ b\end{array}\right] 
  (3\tau | 3(u+n+m\tau),3v)
  &=&
e^{2\pi i na} e^{-6\pi i mv} \bar\Theta_{trap}\left[\begin{array}{c}a \\ b\end{array}\right] 
  (3\tau | 3u,3v)
  \\
  \bar \Theta_{trap}\left[\begin{array}{c}a \\ b\end{array}\right] 
  (3\tau | 3u,3(v+n+m\tau))
  &=&
e^{2\pi i nb} q^{\frac32m^2} e^{-6\pi i m(u-v-1/6)} \bar\Theta_{trap}\left[\begin{array}{c}a \\ b\end{array}\right] 
  (3\tau | 3u,3v)
\ ,\nonumber\eea
and always counts instantons with positive areas, but it does not
extend to a meromorphic section of a line bundle over $\Sigma^2$.

One may wonder about the significance of the extra term in
(\ref{newcorrel}), and more generally, about ambiguities in the
definition of the correlators.  As noted in the introduction, in
general there are contact terms arising from colliding operators
which lead to regularization ambiguities.  In the
closed string sector, contact term ambiguities were no great
deal because they were implicitly fixed by the flat structure of
the moduli space. In the open string sector, no such flat structure
exists for the boundary changing deformations, but we see here that
we may impose other constraints to fix potential ambiguities,
\eg, by insisting on the (quasi-)periodicity of the correlators.

\begin{figure}[t]  
\begin{center}
\includegraphics[width=9.5cm]{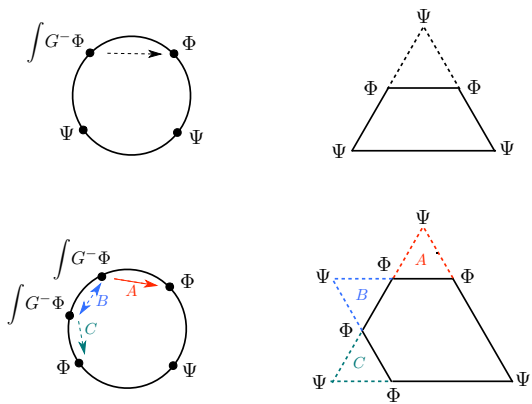}
\caption{At top: the contact term of a trapezoid correlator is given by a fermionic insertion, generated by the collision of two bosonic operators (of which one is integrated). It gets contributions of triangular instantons as sketched on the right hand side.
At bottom: the possible contact terms of the pentagon function are characterized by a parallelogram (B), two kinds of trapezoids (A or C), and a triangle (A plus C).}
\label{fig:contact}
\end{center}
\end{figure}

To see the relevance of contact terms, consider a redefinition of the trapezoidal correlator of the general form
\bea
\ll{contactredef}
 \Theta_{trap}\left[\begin{array}{c}a \\ b\end{array}\right] \ \longrightarrow  \Theta_{trap}\left[\begin{array}{c}a \\ b\end{array}\right] + f \,
 \Theta\left[\begin{array}{c}a+b \\ -3/2\end{array}\right]  
  \eea
for some unspecified, in general moduli-dependent function $f$. It can be given a simple physical interpretation as follows. A contact term arises when
two boundary operators collide and form a single node. The contact
term must thus behave like a three-point function, which by itself is
governed by triangular instantons that are spanned
between the three vertices. Concretely, the only possible
contact term of a trapezoid correlator corresponds to the
triangle that arises when we follow the two converging edges all the way to
their intersection point (see Fig.\ref{fig:contact}, upper part).
From the relative angle this must correspond to a fermionic insertion,
and indeed the charges are such that the collision of two bosonic
operators can generate a contact term $C(\Phi,\Phi)$ of precisely the
charge of a fermion; that is, 
$C(\Phi,\Phi)\sim \Phi(\tau_1)\int_{\tau_1} d\tau_2 G^-\Phi(\tau_2) \sim \Psi(\tau_1)$.

This simple physical picture ties nicely together with results in
the mathematical literature. As was shown in \cite{Polishchukoo},
the origin of the trapezoidal function being
multi-valued and non-modular lies in the existence of a
homomorphism located at the intersection of those two converging
edges. This homomorphism corresponds precisely to the
physical operator content of the contact term, \ie, in our situation,
to~$\Psi$.

Consequently, the parallelogram correlator (\ref{parallelogram}),
for which there are no converging edges, does not suffer from contact
term ambiguities and modular anomalies. This is reflected by the
fact that it can be written in terms of modular theta functions in
the form\footnote{From this we see, similar as for the trapezoid,
that a singularity occurs if either pair of the parallel edges moves
on top of each other, \ie, if $u=0$ or $v=0$.}
$\frac{\Theta'(0)\Theta(u+v)}{\Theta(-u)\Theta(-v)}$, and so
corresponds to a well-defined, meromorphic Jacobi form \cite{PolishchukMA}.

The redefinition (\ref{contactredef}) has a very simple description
also in terms of the effective lagrangian ({\ref{weff}), $\cW_{eff}$.
It just corresponds to the reparametrization, 
$t_a \rightarrow t_a+f_a^{\bar b\bar c}\,\xi_{\bar b}\xi_{\bar c}$, of
the tachyon deformation parameters,
which is compatible with charges and statistics. Inserting this into 
$\cW_{eff}(t_a,\xi_{\bar a})$ trivially reproduces (\ref{contactredef}), \ie:
\bea
\ll{Tredef}
\cT_{ab\bar c\bar d}\ \rarrow\ \cT_{ab\bar c\bar d} +
	   f_{\bar c\bar d}{}^e \Delta_{abe}\ .
\eea	   
This implies, of course, corresponding simultaneous redefinitions
of the higher point correlation functions as well. For example, the
pentagon function will be modified in the following manner:
\bea
\ll{pentaboncont}
\BigP_{a\bar b\bar c\bar e\bar f}\ \rarrow\
\BigP_{a\bar b\bar c\bar e\bar f} +
f_{\bar b\bar c}{}^d \cT_{ad\bar e\bar f}+
f_{\bar e\bar f}{}^d \cT_{ad\bar b\bar c}+
f_{\bar c\bar e}{}^d \cP_{a\bar bd\bar f}+
f_{\bar b\bar c}{}^b f_{\bar e\bar f}{}^c\Delta_{abc}\ .
\eea
The origin of these
terms can be visualized by means of
Fig.\ref{fig:contact}, lower part. On the other hand, the parallelogram
function won't be modified, and this is consistent with the geometric
picture as well.

Whether such a correlated redefinition of all the correlation
functions is compatible with the various $\cA_\infty$ consistency
constraints, may at this point not be entirely obvious, and we will
verify this below by direct computation. However, at tree (disk)
level, its consistency follows also more directly from the structure
of homological perturbation theory (as reviewed in \cite{Lazaroiunm,Kajiuraax,Kajiurasn});
basically, a redefinition of correlators by contact terms with less
legs is consistent if the contact terms satisfy $\cA_\infty$ relations
by themselves.  In this language, it is a homotopy transformation
which by its very definition is compatible with the $\cA_\infty$
structure.

To our knowledge, the corresponding statement has however not been proven
for the $\cA_\infty$ consistency constraints at the {\it quantum level},
and we will verify it for the annulus by direct computation further
below.

\section{Classical $\cA_\infty$ relations}

As it was shown in \cite{HLL}, the topological disk amplitudes that we
determined in the previous sections by instanton counting, should
fulfill certain algebraic equations, the (classical) $\cA_\infty$
relations. These take the form:
\beq 
   \label{Ainfty}
   \sum_{k<l=1}^m (-)^{\td a_1 + \ldots + \td a_k}
   \cF^{0,1}_{a_1\ldots a_k c a_{l+1}\ldots a_m} \rh^{cd} 
   \cF^{0,1}_{d a_{k+1} \ldots a_l} = 0 \tfor m \geq 1\ .
\eeq
Here $\cF^{0,1}_{a_{1} \ldots a_n}$ denotes any one of the disk
amplitudes given in (\ref{threept}) or (\ref{disk}). We suppress the
boundary condition labels for a moment and understand that the $a_i$'s
can take values in the index set $\{\unit,a,\ba,\Omega\}$. Let us
assume for what follows that the `external' fields are only chosen
from the index subset corresponding to boundary changing operators,
\ie, $a_i \in \{a,\ba\}$. The remaining $\cA_\infty$ relations, with
`external' $\Om$ insertions, can be obtained by differentiation with
respect to $u_i$'s. The $\td a_i$'s denote the suspended
$\bbZ_2$-gradings of the boundary fields \cite{HLL}. Here, we have
$\td a = 0$ and $\td\ba = 1$.

From the different types of amplitudes that can appear in our
specific setup it is clear that the relations (\ref{Ainfty}) are
non-trivial only for level $m \in \{4,\ldots,9\}$; for $m=4$ they
just express associativity of the $3$-point correlators. Apart from
the splitting in different levels, $m$, we can further distinguish
two classes of $\cA_\infty$ relations according to the particular
D-brane configuration:

\emph{(i)} The number of D-branes that are involved in the $\cA_\infty$
relation is maximal, \ie, equal to the level $m$. We assume
furthermore that parallel D-branes
are all separated, which means that $u_k\neq u_l$ whenever $\cL_{i_k}$
and $\cL_{i_l}$ are in the same homology class. This is the transversality
condition of ref.~\cite{Polishchukkx}, and we will verify the $\cA_\infty$ 
relations for such transversal D-brane configurations in the next
section.

\emph{(ii)} The number of D-branes is not maximal. This implies that
at least two of the boundary condition labels $i_k$ that appear in the
$\cA_\infty$ relations represent the same D-brane.
Such situations lead to singular correlation functions in view of
degenerate instantons, and we will have to introduce an appropriate
regularization procedure. This configuration is not transversal in the
sense of ref.~\cite{Polishchukkx}, and we will see that it leads to
new results on the $\cA_\infty$ structure on the elliptic curve.

\subsection{$\cA_\infty$ relations for transversal D-brane configurations}

The simplest non-trivial $\cA_\infty$ relation is at level $m=5$ and
involves five D-branes $\cL_i$ for $i=1,\ldots,5$. We pick the field
configuration in (\ref{Ainfty}) to be 
\beq
  \ll{fieldconfm5}
  \{a_1,a_2,a_3,a_4,a_5\} ~~\cong~~
  \{\Psi\bc{1}{2}_{a_1},\Psi\bc{2}{3}_{a_2},\Psi\bc{3}{4}_{a_3},
    \Psi\bc{4}{5}_{a_4},\Phi\bc{5}{1}_{\ba_5}\} \ ,
\eeq
where we assume $u_i\neq u_j$ for $i \neq j$. Naively, for $m=5$ the
sum over $k$ and $l$ in (\ref{Ainfty}) involves five terms. Charge
selection dictates, however, that two of them should involve
$\Om$-insertions, which however are not present because of the assumption of separated branes, \ie, transversality. 
The $\cA_\infty$ relations thus read:  
\bea
  \ll{regularm5}
  \sum_{c=1}^3\cP_{a_1 \bar c a_4 \bar a_5}^{(i_1i_2i_4i_5)}~
  \tria\supers{i_4i_2i_3}_{ca_2a_3}
  +
  \sum_{c=1}^3\cT_{a_1 a_2 \bar c \bar a_5}^{(i_1i_2i_3i_5)}~
  \tria\supers{i_5i_3i_4}_{ca_3a_4}
  +
  \sum_{c=1}^3\tria\supers{i_1i_2i_3}_{a_1a_2c}~
  \cT_{\bar c a_3 a_4 \bar a_5}^{(i_1i_3i_4i_5)}
  = 0 \ .
\eea
Insertion of all the disk amplitudes as given in Section~2 shows
that these relations are indeed satisfied; this transversal situation
has already been studied in \cite{PolishchukMA}. One also readily verifies
that (\ref{regularm5}) is compatible with homotopy reparametrizations
(\ref{Tredef}) that reflect the contact term ambiguity. Indeed,
while the parallelogram function does not allow for a contact term,
the contributions from the trapezoid functions just sum up to:
\bea
\ll{lowestrel}
\sum_{c,e} \Bigl(f_{\bar c\bar a_5}{}^e \tria_{a_1a_2e} \tria_{ca_3a_4}  -
       f_{\bar a_5\bar c}{}^e \tria_{a_1a_2c} \tria_{a_3a_4e}\Bigr)
\eea
and thus cancel for cyclic coefficients, 
$f_{\bar a\bar b}{}^c = f_{\bar b\bar c}{}^a$. The latter condition
ensures that the cyclic invariance of disk amplitudes is
preserved (accordingly, the transformations (\ref{Tredef}) and (\ref{pentaboncont})
satisfying this condition are called \emph{cyclic} homotopies \cite{Polishchukkx,Kajiuraax}).

Let us consider the next level, $m = 6$. A generic $\cA_\infty$ relation
will then involve also pentagon amplitudes. We will come to those
in a moment, but first ask whether
there exist relations at level $m=6$ that do not involve pentagon
amplitudes. Indeed, there are, and they correspond to the field
configuration:
\[
  \{a_1,a_2,a_3,a_4,a_5,a_6\} ~~\cong~~
  \{\Psi\bc{1}{2}_{a_1},\Phi\bc{2}{3}_{\ba_2},
    \Psi\bc{3}{4}_{a_3},\Phi\bc{4}{5}_{\ba_4},
    \Psi\bc{5}{6}_{a_5},\Phi\bc{6}{1}_{\ba_6}\} \ ,
\]
where all $\cL_{i_k}$ for odd $k$ belong to the same homology class. The same
is true for all $\cL_{i_k}$ for even $k$. The $\cA_\infty$ relation takes an
intriguing form that involves only parallelogram amplitudes:
\beq
  \ll{regularm6para}
  \sum_{c=1}^3\cP_{a_1 \bar c a_5 \bar a_6}^{(i_1i_2i_5i_6)}~
  \cP_{c \bar a_2 a_3 \bar a_4}^{(i_5i_2i_3i_4)}
  -
  \sum_{c=1}^3\cP_{a_1 \bar a_2 c \bar a_6}^{(i_1i_2i_3i_6)}~
  \cP_{\bar c a_3 \bar a_4 a_5}^{(i_6i_3i_4i_5)}
  -
  \sum_{c=1}^3\cP_{a_1 \bar a_2 a_3 \bar c}^{(i_1i_2i_3i_4)}
\cP_{c \bar a_4 a_5 \bar a_6}^{(i_1i_4i_5i_6)}
  = 0 \ .
\eeq
Notice that it is manifestly homotopy invariant.
In fact, this relation was interpreted in \cite{PolishchukYB} as an associative
Yang--Baxter equation, for which the $R$-matrix is essentially given by
the parallelogram amplitude. In particular, this link was used to
construct elliptic solutions to the classical Yang--Baxter equation
for $sl_n(\bbC)$. The integer $n$ is an intersection number, in our
situation given by $n = \g(\cL_{i_{2k}},\cL_{i_{2k'+1}})=3$.

There are three further choices for the external fields in equations
(\ref{Ainfty}) for level $m=6$. Two of them, \ie,
\beq
  \nn
  \{a_1,a_2,a_3,a_4,a_5,a_6\} ~~\cong~~
  \begin{array}{l}
  \{\Psi\bc{1}{2}_{a_1},  \Psi\bc{2}{3}_{a_2},
    \Phi\bc{3}{4}_{\ba_3},\Psi\bc{4}{5}_{a_4},
    \Phi\bc{5}{6}_{\ba_5},\Phi\bc{6}{1}_{\ba_6}\}\ , \\[5pt]
  \{\Psi\bc{1}{2}_{a_1},  \Phi\bc{2}{3}_{\ba_2},
    \Psi\bc{3}{4}_{a_3},  \Psi\bc{4}{5}_{a_4},
    \Phi\bc{5}{6}_{\ba_5},\Phi\bc{6}{1}_{\ba_6}\}\ ,
  \end{array}
\eeq
give rise to similar $\cA_\infty$ relations that include pentagon
as well as lower-point amplitudes. We present only one of them here:
\bea
  \ll{regularm6a}
  && \sum_{c=1}^3\De_{a_1 a_2 c}^{(i_1i_2i_3)}~
  \BigP_{\bar c \bar a_3 a_4 \bar a_5 \bar a_6}^{(i_1i_3i_4i_5i_6)}
  +
  \sum_{c=1}^3\cT_{a_1 c \bar a_5 \bar a_6}^{(i_1i_2i_5i_6)}~
  \cP_{\bar c a_2 \bar a_3 a_4}^{(i_5i_2i_3i_4)}
  + \nn \\
  &+&\sum_{c=1}^3\cT_{a_1 a_2 \bar c \bar a_6}^{(i_1i_2i_3i_6)}~
  \cP_{c \bar a_3 a_4 \bar a_5}^{(i_6i_3i_4i_5)}
  -
  \sum_{c=1}^3\cT_{a_1 a_2 \bar a_3 \bar c}^{(i_1i_2i_3i_4)}
\cT_{c a_4 \bar a_5 \bar a_6}^{(i_1i_4i_5i_6)}
  = 0 \ .
\eea
The final $\cA_\infty$ relation at level $m=6$ corresponds to the fields
\[
  \{a_1,a_2,a_3,a_4,a_5,a_6\} ~~\cong~~
  \{\Psi\bc{1}{2}_{a_1},  \Psi\bc{2}{3}_{a_2},
    \Psi\bc{3}{4}_{a_3},  \Phi\bc{4}{5}_{\ba_4},
    \Phi\bc{5}{6}_{\ba_5},\Phi\bc{6}{1}_{\ba_6}\} \ ,
\]
and reads:
\bea
  \ll{regularm6b}
  && \sum_{c=1}^3\De_{a_1 a_2 c}^{(i_1i_2i_3)}~
  \BigP_{\bar c a_3 \bar a_4 \bar a_5 \bar a_6}^{(i_1i_3i_4i_5i_6)}
  + \sum_{c=1}^3
  \BigP_{a_1 \bar c \bar a_4 \bar a_5 \bar a_6}^{(i_1i_2i_4i_5i_6)}~
  \De_{c a_2 a_3}^{(i_4i_2i_3)}
  \nn \\
  &+&
  \sum_{c=1}^3\cT_{a_1 c \bar a_5 \bar a_6}^{(i_1i_2i_5i_6)}~
  \cT_{\bar c a_2 a_3 \bar a_4}^{(i_5i_2i_3i_4)}
  +
  \sum_{c=1}^3\cT_{a_1 a_2 \bar c \bar a_6}^{(i_1i_2i_3i_6)}~
  \cT_{c a_3 \bar a_4 \bar a_5}^{(i_6i_3i_4i_5)}
  = 0 \ .
\eea
Plugging in the instanton sums we verified that all three relations
are indeed satisfied. We can also easily verify invariance of the
level $m=6$ relations (\ref{regularm6a}) and (\ref{regularm6b})
under the combined homotopy transformations (\ref{Tredef}) and
(\ref{pentaboncont}); these map the equations into $\cA_\infty$
relations at lower levels, (\ref{lowestrel}) and (\ref{regularm5}),
that we have already checked to vanish before.

We conclude this section by presenting the $\cA_\infty$ relation at
level $m=7$ that contains hexagon amplitudes (\ref{sixpoint}). For the
fields 
\[
  \{a_1,a_2,a_3,a_4,a_5,a_6,a_7\} ~~\cong~~
  \{\Psi\bc{1}{2}_{a_1},  \Psi\bc{2}{3}_{a_2},
    \Phi\bc{3}{4}_{\ba_3},  \Phi\bc{4}{5}_{\ba_4},
    \Phi\bc{5}{6}_{\ba_5},\Phi\bc{6}{7}_{\ba_6},
    \Phi\bc{7}{1}_{\ba_7}\} \ ,
\]
we get:
\bea
  \ll{regularm7}
  && \sum_{c=1}^3\De_{a_1 a_2 c}^{(i_1i_2i_3)}~
  \cH_{\bar c \bar a_3 \bar a_4 \bar a_5 \bar a_6 \bar a_6}^{(i_1i_3i_4i_5i_6i_7)}
  + \sum_{c=1}^3
  \BigP_{a_1 \bar c \bar a_5 \bar a_6 \bar a_7}^{(i_1i_2i_5i_6i_7)}~
  \cT_{c a_2 \bar a_3 \bar a_4}^{(i_5i_2i_3i_4)} +
  \\ \nn 
  &+& 
  \sum_{c=1}^3\cT_{a_1 c \bar a_6 \bar a_7}^{(i_1i_2i_6i_7)}~
  \BigP_{\bar c a_2 \bar a_3 \bar a_4 \bar a_5}^{(i_6i_2i_3i_4i_5)}
  +
  \sum_{c=1}^3\cT_{a_1 a_2 \bar c \bar a_7}^{(i_1i_2i_3i_7)}~
  \BigP_{c \bar a_3 \bar a_4 \bar a_5 \bar a_6}^{(i_7i_3i_4i_5i_6)}
  -
  \sum_{c=1}^3\cT_{a_1 a_2 \bar a_3 \bar c}^{(i_1i_2i_3i_4)}~
  \BigP_{c \bar a_4 \bar a_5 \bar a_6 \bar a_7}^{(i_1i_4i_5i_6i_7)}
  = 0 \ .
\eea

The other two $\cA_\infty$ relations at level $m=7$ involve only four-
and five-point amplitudes. These relations as well as the missing ones
at level $m=8$ and $9$ can easily be deduced from (\ref{Ainfty}).

\subsection{$\cA_\infty$ relations for non-transversal D-brane configurations}
\ll{sec:nontrans}

Let us consider situation \emph{(ii)} where the number of D-branes is
not maximal, that is, smaller than $m$. Take an $\cA_\infty$ relation
(\ref{Ainfty}) at level $m=5$ for four boundary condition labels, say 
$\cL_{i_1}$, $\cL_{i_2}$, $\cL_{i_3}$ and $\cL_{i_5}$, and the collection
of fields:
\[
  \{a_1,a_2,a_3,a_4,a_5\} ~~\cong~~
  \{\Psi\bc{1}{2}_{a_1},\Psi\bc{2}{3}_{a_2},\Psi\bc{3}{1}_{a_3},
    \Psi\bc{1}{5}_{a_4},\Phi\bc{5}{1}_{\ba_5}\} \ ,
\]
where we assume $u_i\neq u_j$ for $i \neq j$.%
\footnote{Note that this configuration is similar to
  (\ref{fieldconfm5}), only the label $i_4$ was substituted by $i_1$.}

The $\cA_\infty$ relations look similar to (\ref{regularm5}), but now we
do not have a transversal configuration, so that an additional term
appears, where a disk with the fields 
$\langle~\unit\bc{1}{1}\Psi\bc{1}{5}_{a_4}\Phi\bc{5}{1}_{\ba_5}\rangle$ 
bubbles off and gives rise to an amplitude with an insertion of the
boundary condition preserving field $\Om^{i_1i_1}$. Naively we get:
\beq
  \nn
  \sum_{c=1}^3 
  \cP_{a_1 \bar c a_4 \bar a_5}^{(i_1i_2i_1i_5)}
  \tria\supers{i_1i_2i_3}_{ca_2a_3}
  +
  \sum_{c=1}^3 
  \cT_{a_1 a_2 \bar c \bar a_5}^{(i_1i_2i_3i_5)}
  \tria\supers{i_5i_3i_1}_{ca_3a_4}
  +
  \sum_{c=1}^3
  \tria\supers{i_1i_2i_3}_{a_1a_2c}~
  \cT_{\bar c a_3 a_4 \bar a_5}^{(i_1i_3i_1i_5)}
  = - \frac{\rh_{a_4 \bar a_5}}{6\pi i}
  \dl_{u_1}\tria\supers{i_1i_2i_3}_{a_1a_2a_3}
  \ .
\eeq
There is however an important subtlety. The $4$-point functions on the
left-hand side that bear the boundary condition $\cL_{i_1}$ twice can
become singular due to an infinite sum over degenerate
instantons. This happens for instance in the amplitude 
$\cP_{a_1 \bar c a_4 \bar a_5}^{(i_1i_2\td\imath_1i_5)}$ when
opposite sides of a parallelogram collide in the limit $\td u_1 \rarrow u_1$. 
We therefore need to regularize the singular $4$-point functions. We
do this by point-splitting in the following way. Whenever two sides of
a parallelogram (or a trapezoid) bear the \emph{same} D-brane
$\cL_{i_1}$ at position $u_1$, we formally set one of the two sides at
position $\td u_1$ and take the limit $\td u_1 \rarrow u_1$. In order to
track the (formal) $\td u_1$-dependence of the four-point correlators in
the $\cA_\infty$ relation, we introduce the index $\td\imath_1$ and denote the
amplitudes with point-splitting regularization by 
$\widetilde{\cP}_{a_1 \bar c a_4 \bar a_5}^{(i_1i_2\td\imath_1i_5)}$ and 
$\widetilde{\cT}_{\bar c a_3 a_4 \bar a_5}^{(i_1i_3\td\imath_1i_5)}$. 
The $\cA_\infty$ relations then become:  
\beq
  \ll{degenm5}
  \lim_{\td u_1\rarrow u_1} \sum_{c=1}^3  \left(
  \widetilde{\cP}_{a_1 \bar c a_4 \bar a_5}^{(i_1i_2\td\imath_1i_5)}
  \tria\supers{i_1i_2i_3}_{ca_2a_3}
  +
  \cT_{a_1 a_2 \bar c \bar a_5}^{(i_1i_2i_3i_5)}
  \tria\supers{i_5i_3i_1}_{ca_3a_4}
  +
  \tria\supers{i_1i_2i_3}_{a_1a_2c}~
  \widetilde{\cT}_{\bar c a_3 a_4 \bar a_5}^{(i_1i_3\td\imath_1i_5)}\right)
  = - \frac{\rh_{a_4 \bar a_5}}{6\pi i}
  \dl_{u_1}\tria\supers{i_1i_2i_3}_{a_1a_2a_3}
  \ .
\eeq
Using the transversal $\cA_\infty$ relation (\ref{regularm5}) with $i_4
= \td\imath_1$ we see that the singularities of the right-hand side
of (\ref{degenm5}) mutually cancel. So we can safely take the limit 
$\td u_1 \rarrow u_1$ and verify the relation.

Analogously, there is a non-transversal version of the level $m=6$
relation (\ref{regularm6para}) that includes only parallelograms. For this, let
us consider the field configuration:
\[
  \{a_1,a_2,a_3,a_4,a_5,a_6\} ~~\cong~~
  \{\Psi\bc{1}{2}_{a_1},\Phi\bc{2}{1}_{\ba_2},
    \Psi\bc{1}{4}_{a_3},\Phi\bc{4}{5}_{\ba_4},
    \Psi\bc{5}{6}_{a_5},\Phi\bc{6}{1}_{\ba_6}\} \ ,
\]
Following the same regularization procedure as above we obtain the
$\cA_\infty$ relations:
\beq
  \nn
  \lim_{\td u_1\rarrow u_1} \sum_{c=1}^3  \left(
  \cP_{a_1 \bar c a_5 \bar a_6}^{(i_1i_2i_5i_6)}~
  \cP_{c \bar a_2 a_3 \bar a_4}^{(i_5i_2i_1i_4)}
  \!-\!
  \widetilde{\cP}_{a_1 \bar a_2 c \bar a_6}^{(i_1i_2\td\imath_1i_6)}~
  \cP_{\bar c a_3 \bar a_4 a_5}^{(i_6i_1i_4i_5)}
  \!-\!
  \widetilde{\cP}_{a_1 \bar a_2 a_3 \bar c}^{(i_1i_2\td\imath_1i_4)}
  \cP_{c \bar a_4 a_5 \bar a_6}^{(i_1i_4i_5i_6)} \right)
  \!=\!
  \frac{\rh_{a_1 \bar a_2}}{6\pi i} 
  \dl_{u_1} \cP_{a_3 \bar a_4 a_5 \bar a_6}^{(i_1i_4i_5i_6)}\ ,
\eeq
which are manifestly homotopy invariant.

We refrain from going through the list of remaining non-transversal
$\cA_\infty$ relations here, the general picture should be clear from the
cases that we presented so far.

\section{Quantum $\cA_\infty$ relations: the annulus}

In order to get a handle on higher genus topological string amplitudes
with multiple boundary components, we can take advantage of the quantum
$\cA_\infty$ relations of \cite{Herbstkt} which follow from
factorizations of higher genus amplitudes. For the elliptic curve, the
charges of the boundary operators are such that the only non-vanishing
open topological string amplitudes beyond tree level appear at one
loop, \ie, for annulus world-sheets.
The charge selection rule for the topological amplitude is quite
restrictive and implies for the D-brane geometry at hand that there
are only two non-trivial factorization relations for the annulus. All
others can be obtained by differentiation with respect to the
boundary moduli $u_i$. 

\begin{figure}[t]  
\begin{center}
\includegraphics[width=14cm]{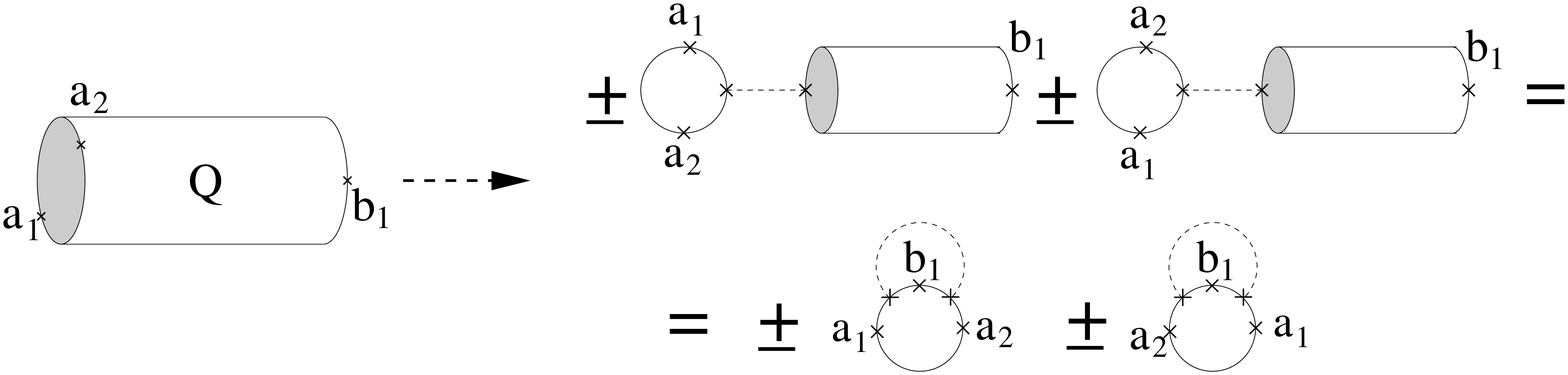}
\caption{A graphical representation of the  quantum $\cA_\infty$ relation, given in
  (\ref{annulusfact}).}
\label{fig:ann1}
\end{center}
\end{figure}

The first non-trivial relation is diagrammatically depicted in 
Fig.\ref{fig:ann1}. Explicitly, denoting by $\cF^{0,1}$ and
$\cF^{0,2}$ the generic topological string amplitude on the disk and 
annulus, respectively, this quantum $\cA_\infty$ relation takes the
form \cite{Herbstkt}: 
\bea
  \ll{annulusfact}
  && \sum_{c,d}\bigl((-)^{\td a_1 + \td d \td a_2}~  
     \cF^{0,1}_{a_1 c a_2} ~\rh^{cd} ~\cF^{0,2}_{d|~b_1} +
     (-)^{\td a_1 + \td a_2}~         
     \cF^{0,1}_{a_1 a_2 c} ~\rh^{cd} ~\cF^{0,2}_{d|~b_1} \bigr) 
  \nn\\[5pt]
  &=&
  \sum_{c,d} \bigl(
    (-)^{\td a_1+\td b_1(\td d+\td a_2)} 
    \rh^{cd} \cF^{0,1}_{a_1 c b_1 d a_2} + 
    (-)^{\td a_1 + \td a_2+\td b_1 \td d} 
    \rh^{cd} \cF^{0,1}_{a_1 a_2 c b_1 d}
  \bigr)  \ .
\eea
Here, the labels $a_i$ subsume both field and boundary
condition labels, and $\rh^{ab}$ denotes the inner product of the boundary
fields (\ref{metric}). Without loss of generality we pick the
following choice of boundary operators in (\ref{annulusfact}):
\[ a_1 \sim \Psi\fla\bc{1}{2} \qquad
   a_2 \sim \Phi\flbb\bc{2}{1} \qquad
   b_1 \sim \Om\bc{3}{3} \ .
\]
The reason why we insert the boundary preserving modulus $\Omega\bc{3}{3}$
on the second boundary, is that the factorization problem of open
higher genus amplitudes is well-defined only if there is at least
one topological observable on each boundary component \cite{Herbstkt}.
Otherwise the boundary without insertions bubbles off in the closed
string channel and gives rise to a non-stable disk amplitude. This
corresponds to a noncompact direction in the moduli space of the Riemann
surface and, as we will discuss below, indicates divergences in the
amplitudes.

From the field configuration it is clear
that $\cL_{i_1}$ and $\cL_{i_2}$ wrap different homology classes.
Introducing the following notation for annulus correlators:
\beq
  \cA_{\Om\ldots|\Om\ldots}^{(i_1i_1\ldots|i_2i_2\ldots)} \ =\ 
      \bra \Om\bc{1}{1}\ldots |\Om\bc{2}{2}\ldots
      \ket_{ann}\ ,
\eeq
the annulus $\cA_\infty$ relation (\ref{annulusfact}) then
simplifies to 
\bea
  \ll{annulushere}
  \rh_{a \bar b}\left(
  \cA^{(\ial_1\ial_1|\ial_3\ial_3)}_{\Om|~\Om}  
  -
  \cA^{(\ial_2\ial_2|\ial_3\ial_3)}_{\Om|~\Om}\right)  &=&
  \sum_{c=1}^3 \bigl( 
     \cT^{\ial_1\ial_2\ial_3\ial_3\ial_2}_{a c \Om \bar c \bar b} - 
     \cP^{\ial_1\ial_2\ial_1\ial_3\ial_3}_{a \bar b c \Om \bar c} 
  \bigr) + \\
  \nn &+& \sum_{c=1}^3 \bigl( 
     \cP^{\ial_1\ial_2\ial_3\ial_3\ial_2}_{a \bar c \Om c \bar b} - 
     \cT^{\ial_3\ial_1\ial_2\ial_1\ial_3}_{c a \bar b \bar c \Om} 
  \bigr)  \ .
\eea
Note that this equation is compatible with homotopy transformations,
\ie, contact term redefinitions of the trapezoid
amplitudes as given in (\ref{Tredef}). Specifically, a homotopy
transformation adds
\[ \partial_{u_3} \Bigl( 
   \sum_{c,e} (f_{\bar c\bar b}{}^e \Delta_{ace}- 
               f_{\bar b\bar c}{}^e \Delta_{cae}) 
   \Bigr)
\]
to the right-hand side of (\ref{annulushere}), which vanishes for
cyclic coefficients, \ie, $f_{\bar a\bar b}{}^c = f_{\bar b\bar c}{}^a$.

\vskip .5cm

For the third homology class $i_3$ we have two choices: \emph{(i)}
either $\cL_{i_3}$ wraps a homology class different from both, $\cL_{i_1}$
and $\cL_{i_2}$, or \emph{(ii)} it wraps the same
class as $\cL_{i_1}$ (or $\cL_{i_2}$).  We will now discuss these
two cases separately: \vskip2mm \emph{(i)} When all three D-branes
wrap different homology classes, the quantum $\cA_\infty$ relations
become:
\bea
  \ll{annulusI}
  \rh_{a \bar b}\left(
  \dl_{u_3}\cA^{(\ial_1\ial_1|\ial_3)}_{\Om|~\cdot}  
  -
  \dl_{u_3}\cA^{(\ial_2\ial_2|\ial_3)}_{\Om|~\cdot}\right)  &=&
  \sum_{c=1}^3 \bigl( 
     \dl_{u_3}\cT^{\ial_1\ial_2\ial_3\ial_2}_{a c \bar c \bar b} 
   - \dl_{u_3}\cT^{\ial_3\ial_1\ial_2\ial_1}_{c a \bar b \bar c} 
  \bigr) = \\ &=& 
  \nn \sum_{c=1}^3 \rh_{a \bar b} \bigl( 
     \dl_{u_3}\cT^{\ial_1\ial_2\ial_3\ial_2}_{a c \bar c \bar a} 
   - \dl_{u_3}\cT^{\ial_3\ial_1\ial_2\ial_1}_{c a \bar a \bar c} 
  \bigr) = 0 \ .
\eea
Here we wrote the field insertion $\Om^{(\ial_3\ial_3)}$ in terms of
derivatives with respect to $u_3$. In the second line we used the Kronecker
deltas from (\ref{trapezoid}) and (\ref{parallelogram}). The last step follows
from explicitly inserting the trapezoid amplitudes (\ref{trapezoid}).
By considering analogous relations for other choices for the $\cL_i$'s
it follows readily that
\[ \cA^{(\ial_1\ial_1|\ial_2\ial_2)}_{\Om|~\Om} = 
   \cA^{(\ial_2\ial_2|\ial_3\ial_3)}_{\Om|~\Om} = 
   \cA^{(\ial_3\ial_3|\ial_1\ial_1)}_{\Om|~\Om} = 
   \frac1{(6\pi i)^2}{f_\cA(\tau)} \ ,
\]
for some $f_\cA(\tau)$. Since this function is $u_i$-independent, it cannot
be an instanton series and so must be simple. This fact is also clear
from the geometric picture, \ie, it is not possible to span an annulus
between non-parallel D-branes. In principle, we could
determine $f_\cA(\tau)$ via imposing modular invariance, but we will identify
it below by comparison with a known result.

\vskip2mm
\emph{(ii)} If, say, $\cL_{i_1}$ and $\cL_{i_3}$ wrap the same
homology class, then the annulus factorization condition
(\ref{annulushere}) simplifies to 

\bea
  \ll{annulusII}
  \rh_{a \bar b}\left(
  \dl_{u_3}\cA^{(\ial_1\ial_1|\ial_3)}_{\Om|~\cdot}  
  - \frac{f_\cA(\ta)}{6\pi i}
  \right)  &=&
  \sum_{c=1}^3 \rh_{a \bar b} 
  \dl_{u_3} \cP^{\ial_1\ial_2\ial_3\ial_2}_{a \bar c c \bar a} 
  \ ,
\eea
which is manifestly homotopy invariant. The function 
$\dl_{u_3}\cA^{(\ial_2\ial_2|\ial_3)}_{\Om|~\cdot}=1/(6\pi i)f_\cA(\ta)$ 
appears here because $\cL_{i_2}$ and $\cL_{i_3}$ wrap different
homology classes. 
  
Notice that in the disk correlators on the right-hand side of
(\ref{annulusII}), one pair of parallel sides of the parallelograms
corresponds to the same D-brane, $\cL_{i_2}$; we thus encounter a
non-transversal configuration and need to regularize the correlators
in order to evaluate the sum. Its divergent part is, however,
$u_3$-independent and gets 
annihilated by the $u_3$-derivative, so that the right-hand side
of (\ref{annulusII}) is well-defined. Had we not inserted the
boundary modulus $\Omega\bc{3}{3}$ in the first place and thus had
considered the integrated version of (\ref{annulusII}), the
(non-cancelling) divergent pieces of the parallelogram correlators
would not have been killed; this divergence reflects the non-stable
closed string degeneration channel where a disk with a ``bare''
boundary bubbles off \cite{Herbstkt}.%
\footnote{In a sense, infinitely many degenerate
  parallelogram instantons on the right hand side of (\ref{annulusII})
  conspire to reproduce the singularity arising from a non-stable
  degeneration, \ie, from not having fixed
  the isometries of a disk that pinches off.}

\begin{figure}[t]  
\begin{center}
\includegraphics[width=4.5cm]{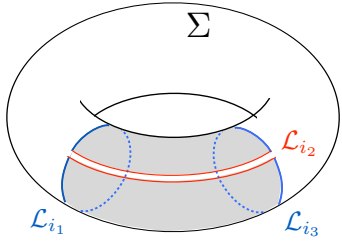}
\caption{Shown is how annulus instantons can be obtained from
disk instantons via the fusion of boundary components. This is what
underlies geometrically the quantum $\cA_\infty$ relation (\ref{annulusII}).
The divergence arising from coinciding boundary components disappears
if we consider a well-defined stable degeneration limit, by choosing
suitable operator insertions.}
\label{fig:annulusfig}
\end{center}
\end{figure}
Since $\cL_1$ and $\cL_3$ are parallel, we expect world-sheet
instantons with the topology of an annulus to contribute to
$\cA^{(\ial_1\ial_1|\ial_3)}_{\Om|~\cdot}$.  Indeed, if we insert in
(\ref{annulusII}) the parallelogram series
$\cP\sim\Theta_{para}
$ 
as given in (\ref{parallelogram}), we get
\bea
 \label{diskannulusmap}
  \dl_{u_3}\cA^{(\ial_1\ial_1|\ial_3)}_{\Om|~\cdot} &=& 
  \frac{f_\cA(\ta)}{6\pi i} + \sum_{c=1}^{3} 
 \,  \dl_{u_3} \Theta_{para}\left[\!\!
    \begin{array}{c} 3 \\[0pt] [a-c]_3\end{array}
    \!\!\right] (3\tau | 0,3(u_1-u_3)) \nn \\  &=&
  \frac{f_\cA(\ta)}{6\pi i} + \sum_{b=1}^{3} 
 \,  \dl_{u_3} \Theta_{para}\left[\!\!
    \begin{array}{c} 3 \\[0pt] b \end{array}
    \!\!\right] (3\tau | 0,3(u_1-u_3)) \nn \\  &=&
  \frac{f_\cA(\ta)}{6\pi i} + \sum_{b=1}^{3} 
\,  \dl_{u_3}\!\! \sum_{n\neq -1,m \in \bbZ}^{\mathrm{indef.}}
  q^{(n+1)(b + 3m)} 
  e^{6\pi i (n+1) (u_1-u_3)} \nn \\  &=&
  \frac{f_\cA(\ta)}{6\pi i} + 
  \dl_{u_3}\!\! \sum_{n\neq 0,m \in \bbZ}^{\mathrm{indef.}}
  q^{nm} e^{6\pi i n (u_1-u_3)} \nn 
  \ .
\eea
The interpretation of this series is obvious: it should describe the second
derivative of the annulus instanton sum for parallel D-branes, $\cL_{i_k}$ and $\cL_{i_l}$. Up to an integration constant, we can read it off as follows:
\beq
  \ll{anninst}
  \cA^{(i_k|i_l)}_{\cdot|\cdot} =-\frac12 f_A(\tau)(u_k-u_l)^2+
  \sum_{n\neq 0,m \in \bbZ}^{\mathrm{indef.}}
  \frac{1}{n}q^{nm} e^{6\pi i n (u_k-u_l)} \ .
\eeq
This indeed coincides with the result given in \cite{mirrbook} for the
annulus partition function in the holomorphic limit, $\cF^{(0,2)}$, 
provided we identify:
$$
f_A(\tau)\ =\ {\rm const.}\tau\ .
$$
This term is thus a remnant of the holomorphic anomaly of the annulus amplitude.

Note that the annulus instanton series (\ref{anninst}) was obtained,
via the $\cA_\infty$ factorization relation (\ref{annulusfact}), from
disk correlators, which by themselves were determined by counting disk 
instantons. Geometrically, this implies that the annulus instanton
contributions can be patched together 
in terms of disk instantons, in a similar spirit as the $\cA_\infty$
relations on the disk imply the patching up of higher $N$-gons
in terms of  smaller $N$-gons. Indeed, there
is a very simple geometrical picture that describes the instanton
geometry underlying eq.~(\ref{annulusII}), and this is
schematically shown in Fig.\ref{fig:annulusfig}.

The second non-trivial annulus factorization condition on the torus is
schematically depicted in Fig.\ref{fig:ann2}. Before we present the
explicit quantum $\cA_\infty$ relation, let us pick the boundary
operators to be:
\[ a_1 \sim \Phi_{\bar a_1}\bc{1}{2} \qquad
   a_2 \sim \Phi_{\bar a_2}\bc{2}{3} \qquad
   a_2 \sim \Phi_{\bar a_3}\bc{3}{1} \qquad
   b_1 \sim \Om\bc{4}{4} \ .
\]
From the charge selection rule it is quite straightforward to see
that all factorization channels, where disks bubble off from the
annulus, must vanish. Another immediate consequence of the field
configuration at hand is that $\cL_{i_k}$ for $k=1,2,3$ must wrap
three different homology classes. Let us choose the fourth D-brane
$\cL_{i_4}$ to be parallel to, say, $\cL_{i_1}$. 
\begin{figure}[t]  
\begin{center}
\includegraphics[width=16cm]{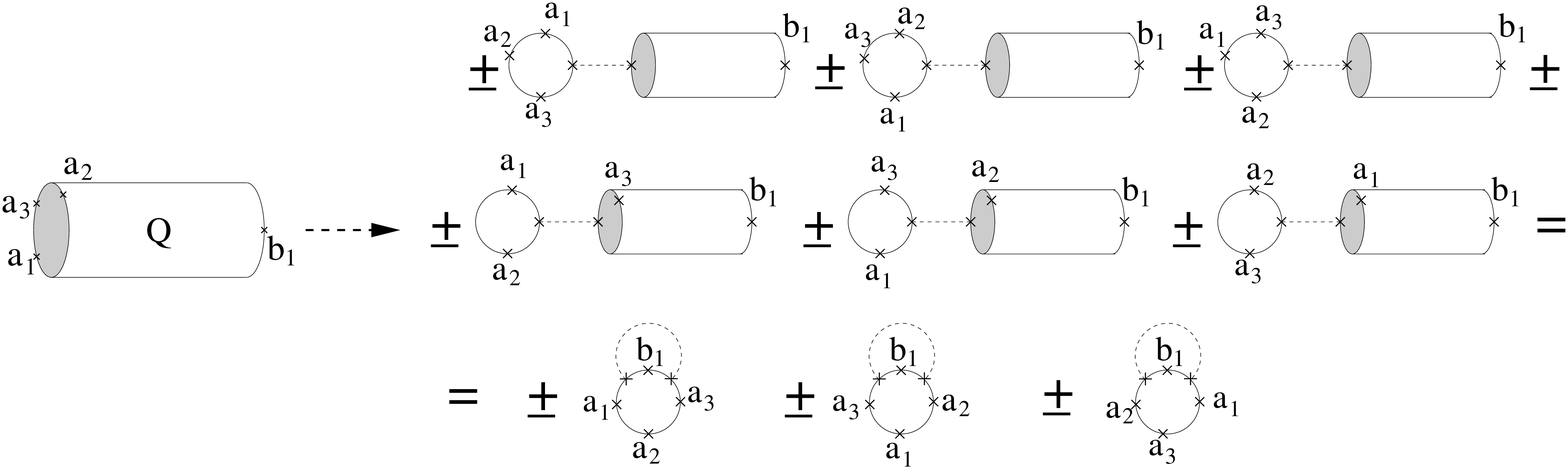}
\caption{A graphical realization of the quantum $\cA_\infty$ relation
  that leads to (\ref{annulusfact3}). In our situation only two diagrams
  on the right-hand side are associated with non-vanishing amplitudes.}
\label{fig:ann2}
\end{center}
\end{figure}

Then the algebraic relation associated to
Fig.\ref{fig:ann2} simplifies considerably and we get the following
constraint on pentagon correlation functions:
\beq
  \ll{annulusfact3}
  \sum_{c} \left(
    \dl_{u_4}\BigP^{(i_1i_2i_3i_4i_3i_1)}_{\ba_1 \ba_2 \bar c c \ba_3}
    - \dl_{u_4}\BigP^{(i_1i_2i_4i_2i_3i_1)}_{\ba_1 c \bar c \ba_2 \ba_3}
  \right) = 0 \ ,
\eeq
where we have already substituted the proper labels for the
operators in question. 
Inserting the instanton sum (\ref{fivepoint}) for the five-point
amplitudes we learn that relation (\ref{annulusfact3}) is indeed
satisfied. Moreover, it is easily shown to
be homotopy invariant for cyclic coefficients $f_{\bar a\bar b}{}^c$,
provided we use the annulus relations (\ref{annulusI}) and (\ref{annulusII}).

\vfill \eject

\paragraph{\bf Acknowledgments}

We thank the KITP Santa Barbara, where part of this work was carried
out, for kind hospitality. This part of the work was supported by
the National Science Foundation under Grant No. PHY99-07949.  Moreover
we thank Ezra Getzler, Suresh Govindarajan, Hans Jockers and Nick
Warner for instructive discussions.

\goodbreak
\appendix
\noindent
\section{Heat equation from the Cardy relation?}
\label{sec:appendix}
\vskip 1em

In \cite{HLL} another kind of annulus factorization was discussed,
in the spirit of the topological ``Cardy-constraint''.  It essentially
equates open and closed string channels of the annulus diagram, and
reads, in the notation introduced above:
\bea
\ll{Cardy}
  &&\partial_i \cF^0_{a_0 \ldots a_n} \eta^{ij}
  \partial_j \cF^0_{b_0 \ldots b_m} = \\[5pt]
  &=&\hspace*{-9mm}
  \sum\limits_{\tiny \begin{array}{c}0\leq n_1 \leq n_2 \leq n \\
      0\leq m_1 \leq m_2 \leq m \end{array}}\hspace*{-7mm}
  (-1)^{(\td c_1+\td a_0)(\td c_2+\td b_0)+\td c_1 +\td c_2}\!
  \rho^{c_1d_1} \rho^{c_2d_2}
  \cF^{0,1}_{a_0 \ldots a_{n_1} d_1 b_{m_1\!\!+\!1} \ldots b_{m_2} c_2 a_{n_2\!+\!1}
  \ldots a_{n}}~
  \cF^{0,1}_{b_0 \ldots b_{m_1} c_1 a_{n_1\!\!+\!1} \ldots a_{n_2} d_2 b_{m_2\!+\!1}
  \ldots b_{m}}
  \;,\nonumber
\eea
where $\cF^0$ and $\eta^{ij}$ 
are prepotential and metric of the closed string sector, respectively.
However, as already pointed out in \cite{HLL}, while the topological
Cardy condition is one of the basic axioms of boundary TFT
\cite{MooreSegal,Moore,CIL}, it needs to apply only to correlators
without integrated insertions (for example, to the topological
intersection amplitude).  Whenever there are integrated insertions,
so that we deal with topological strings rather than with TFT, the
proof of the independence of the correlator of the 
annulus metric does not necessarily go through,  and so it may be
somewhat unclear whether the Cardy constraint should be imposed in
such a situation or not.

In fact, there are results pointing in either way: it was shown in
\cite{Brunnerdc} that for boundary LG models with arbitrarily
deformed univariate superpotentials, the Cardy constraint is
satisfied. Moreover, it was shown in \cite{HLL} that the Cardy
constraint can be imposed on the correlators of the $A$-series of
boundary minimal models and this does in fact lead precisely to the
correct effective action \cite{Herbstzm}. On the other hand,
recent results \cite{KnappOmer} indicate that the Cardy condition
cannot be consistently imposed on correlators of minimal models
other than the $A$-series.

One of the original motivations for our present work was to see
whether the Cardy condition can be imposed on correlators on the
elliptic curve. This question appeared to be potentially interesting
also from a different perspective, namely from the heat equation
that is satisfied by the three-point correlators:\footnote{This is
one of the defining equations for the ordinary theta-functions;
analogous equations hold for the higher point correlators which are
given in terms of indefinite theta-functions.}
\bea
\ll{heateq}
\left(\frac\partial{\partial\tau}+\frac i{12\pi}\frac{\partial^2}{\partial {u_i}^2}\,\right)\,\Delta_{abc}(\tau|u_i)\ =\ 0\ .
\eea
The question is, whether, as discussed in \cite{Brunnermt},
this equation simply reflects the underlying operator algebra of
the model, or whether there is a deeper reason behind it -- such
as some form of background (in-)dependence. From this point of view,
one may interpret (\ref{heateq}) as telling how a change of open
string background ($u_i$) can be compensated by a change of closed
string background ($\tau$), or vice versa.

Our initial observation was that the heat equation (\ref{heateq})
may be linked to the Cardy condition (\ref{Cardy}) as follows.
Consider relation (\ref{Cardy}) with $n=2$, $m=0$, and
\[
  \{a_0,a_1,a_2,b_0\} \cong
  \{\Psi\bc{1}{2}_a,\Psi\bc{2}{3}_b,\Psi\bc{3}{1}_c,\Omega\bc{1}{1}\}\ .
\]
Note that the boundary condition $\cL_{i_1}$ appears on both sides of
the annulus; we encounter a non-transversal configuration as in
Section~\ref{sec:nontrans}, which will require some regularization. 
Taking everything together, the Cardy condition (\ref{Cardy})
then reduces to the following equation:
\bea
\ll{isCardyHeat?}
-\frac2{3\pi i}\frac\partial{\partial\tau}\Delta_{abc}(u_1+u_2+u_3)
\! &\!+\!&\!2\,
\Delta_{abc\Omega\Omega}(u_1+u_2+u_3) \\
&=& \sum_{e,f}\pm\widetilde{\cT}_{ab\bar d\bar e}(u_1+u_2+u_3,u_1-\td u_1)\,\Delta_{dce\Omega}(u_2+u_3+u_1)\ ,\nonumber
\eea
with the understanding that we need to take the limit $\td u_1\rightarrow
u_1$.  Converting the $\Omega$-insertions into $u$-derivatives, we
see that the LHS of this equation indeed coincides with the heat
equation, provided the RHS vanishes. By inserting the explicit
expressions for the correlators, it however turns out that the RHS
does not vanish. One may be tempted to make use of homotopy
transformations of the form (\ref{Tredef}) to remove it, but by
their nature as theta-functions they cannot cancel the singularity
of the trapezoidal instanton sum, \ie, the Appell function. Thus,
by presenting a counter-example, we conclude that the Cardy constraint
(\ref{Cardy}) does not hold for the elliptic curve, and specifically
that the heat equation (\ref{heateq}) is not implied by it.




\catcode`\@=11
\def\mref#1{\ifx\und@fined#1{Need to supply reference \string#1.}\else #1 \fi}
\catcode`\@=12

\def\HLL{
  M.~Herbst, C.~I.~Lazaroiu and W.~Lerche,
  ``Superpotentials, $\cA_\infty$ relations and WDVV equations for open
  topological strings,''
  JHEP {\bf 0502}, 071 (2005)
  [arXiv:hep-th/0402110].}

 \def\DVV{ R.~Dijkgraaf, H.~Verlinde and E.~Verlinde,
 ``Topological Strings in D $<$ 1,'' Nucl.\ Phys.\ B {\bf 352}, 59 (1991).}

\def\kontsevich{
M.~Kontsevich, ``Homological algebra of mirror symmetry,'' in {\em Proceedings
  of the International Congress of Mathematicians, Vol.\ 1, 2 (Z\"urich,
  1994)}, (Basel), pp.~120--139, Birkh\"auser, 1995.}
  
 \def\mirrbook{K.~Hori, S.~Katz, A.~Klemm, R.~Pandharipande, 
 R.~Thomas, C.~Vafa, R.~Vakil and E.~Zaslow,  ``Mirror Symmetry'', 
 Clay Mathematics Monographs V 1, American Mathematical Society, July 2003.} 

\def\Blumenhagenmu{
  R.~Blumenhagen, M.~Cvetic, P.~Langacker and G.~Shiu,
  ``Toward realistic intersecting D-brane models,''
 [arXiv:hep-th/0502005].}

 \def\MooreSegal{G. Moore and G. Segal, unpublished; 
 see http://online.kitp.ucsb.edu/online/mp01/}
 
 \def\Moore{G.~W.~Moore,
 ``Some comments on branes, $G$-flux, and $K$-theory,''
 Int.\ J.\ Mod.\ Phys.\ A {\bf 16}, 936 (2001) [arXiv:hep-th/0012007].}
 
  \def\CIL{C.~I.~Lazaroiu,  
 ``On the structure of open-closed topological field theory in two dimensions,''
 Nucl.\ Phys.\ B {\bf 603}, 497 (2001) [arXiv:hep-th/0010269].}

\def\Lazaroiunm{
  C.~I.~Lazaroiu,
  ``String field theory and brane superpotentials,''
  JHEP {\bf 0110}, 018 (2001)
  [arXiv:hep-th/0107162].}

\def\Cremadesqj{
  D.~Cremades, L.~E.~Ibanez and F.~Marchesano,
  ``Yukawa couplings in intersecting D-brane models,''
  JHEP {\bf 0307}, 038 (2003)
  [arXiv:hep-th/0302105].}
  
\def\Cremadeswa{
  D.~Cremades, L.~E.~Ibanez and F.~Marchesano,
  ``Computing Yukawa couplings from magnetized extra dimensions,''
  JHEP {\bf 0405}, 079 (2004)
  [arXiv:hep-th/0404229].}

\def\Govindarajanim{
  S.~Govindarajan, H.~Jockers, W.~Lerche and N.~P.~Warner,
  ``Tachyon condensation on the elliptic curve,''
  [arXiv:hep-th/0512208].}
  
\def\Brunnermt{
  I.~Brunner, M.~Herbst, W.~Lerche and J.~Walcher,
  ``Matrix factorizations and mirror symmetry: The cubic curve,''
 [arXiv:hep-th/0408243].}
  
\def\Polishchukkx{
  A.~Polishchuk,
  ``$\cA_{\infty}$-structures on an elliptic curve,''
  Commun.\ Math.\ Phys.\  {\bf 247}, 527 (2004)
  [arXiv:math.ag/0001048].}

\def\Polishchukoo{
  A.~Polishchuk,
  ``Indefinite theta series of signature (1,1) from the point of view of homological mirror symmetry,''
   [arXiv:math.ag/0003076].}

\def\Lazaroiukm{
  C.~I.~Lazaroiu,
  ``Non-commutative moduli spaces of topological D-branes,''
  [arXiv:hep-th/0511049].}

\def\Polishchukdb{
  A.~Polishchuk and E.~Zaslow,
  ``Categorical mirror symmetry: The Elliptic curve,''
  Adv.\ Theor.\ Math.\ Phys.\  {\bf 2}, 443 (1998)
  [arXiv:math.ag/9801119].}

\def\PolishchukYB{
  A.~Polishchuk,
  ``Classical Yang-Baxter equation and the $\cA_{\infty}$-constraint,''
  [arXiv:math.AG/0008156].}

\def\PolishchukAP{
  A.~Polishchuk,
  ``M. P. Appell's function and vector bundles of rank 2 on elliptic curve,''
  [arXiv:math.AG/9810084].}

\def\PolishchukMA{
  A.~Polishchuk,
  ``Massey and Fukaya products on elliptic curves,''
  [arXiv:math.AG/9803017].}
  
\def\PolishchukHMS{
  A.~Polishchuk,
  ``Homological mirror symmetry with higher products,''
  [arXiv:math.AG/9901025].}

  \def\horiinf{K.~Hori and J.~Walcher, 
``F-term equations near Gepner points,''  
JHEP {\bf 0501} (2005) 008 
[hep-th/0404196].}

\def\Brunnerdc{
  I.~Brunner, M.~Herbst, W.~Lerche and B.~Scheuner,
  ``Landau-Ginzburg realization of open string TFT,''
 [arXiv:hep-th/0305133].}
  
\def\Herbstzm{
  M.~Herbst, C.~I.~Lazaroiu and W.~Lerche,
  ``D-brane effective action and tachyon condensation in topological  minimal
  models,''
  JHEP {\bf 0503}, 078 (2005)
  [arXiv:hep-th/0405138].}
  
  \def\KnappOmer{
  J.~Knapp and H.~Omer, to appear.}

\def\Kajiurasn{
  H.~Kajiura and J.~Stasheff,
  ``Open-closed homotopy algebra in mathematical physics,''
  [arXiv:hep-th/0510118].}
  
\def\Kajiuraax{
  H.~Kajiura,
  ``Noncommutative homotopy algebras associated with open strings,''
  [arXiv:math.qa/0306332].}
  
\def\Herbstkt{
  M.~Herbst,
  ``Quantum A-infinity structures for open-closed topological strings,''
  [arXiv:hep-th/0602018].}
  
\def\Aganagicgs{
  M.~Aganagic and C.~Vafa,
  ``Mirror symmetry, D-branes and counting holomorphic discs,''
 [arXiv:hep-th/0012041].}
  
\def\Aganagicnx{
  M.~Aganagic, A.~Klemm and C.~Vafa,
  ``Disk instantons, mirror symmetry and the duality web,''
  Z.\ Naturforsch.\ A {\bf 57}, 1 (2002)
  [arXiv:hep-th/0105045].}
  
\def\Lossev{
  A.~Lossev,
  ``Descendants constructed from matter field and K. Saito higher residue
  pairing in Landau-Ginzburg theories coupled to topological gravity,''
TPI-MINN-92-40T.}


%

\end{document}